\documentclass{emulateapj}
\usepackage{psfig}
\def\msol{\hbox{$\rm\thinspace M_{\odot}$}} \def\etal{{\it et
al.\thinspace}} \def\eg{{\it e.g.\ }}

\def\p3m{P${}^3$M} \def\ap3m{AP${}^3$M} \def\-{{\em{---}}}
 \def\msun{{M_\odot}}

\newcommand{\be}{\begin{equation}} \newcommand{\ba}{\begin{eqnarray}}
\newcommand{\ee}{\end{equation}} \newcommand{\ea}{\end{eqnarray}}
\lefthead{Scannapieco, Thacker, \& Couchman}  \righthead{}

\begin{document}

\def\lesssim{\mathrel{\hbox{\rlap{\hbox{\lower4pt\hbox{$\sim$}}}\hbox{$<$}}}}
\def\gtrsim{\mathrel{\hbox{\rlap{\hbox{\lower4pt\hbox{$\sim$}}}\hbox{$>$}}}}
\def\ApJ{{\em Astrophys.\ J.\ }} \def\AJ{{\em Astron.\ J.\ }}
\def\MNRAS{{\em Mon.\ Not.\ R.\ Astron.\ Soc.\ }}
\def\CIV{\hbox{C$\scriptstyle\rm IV\ $}}\

\title{Measuring AGN Feedback with the Sunyaev-Zel'dovich Effect}

\author{Evan Scannapieco\altaffilmark{1},  Robert
J. Thacker\altaffilmark{2,3,4},  and
H. M. P. Couchman\altaffilmark{5}}
\altaffiltext{1}{School of Earth and Space Exploration,  Arizona State University, PO
Box 871404, Tempe, AZ, 85287-1404.}  \altaffiltext{2}{Department of
Physics, Engineering Physics and  Astronomy, Queen's University,
Kingston, Ontario, K7L 3N6, Canada.}  \altaffiltext{3}{New Address:
Department of Astronomy and Physics, Saint  Mary's  University,
Halifax, Nova Scotia, B3H 3C3, Canada.}  \altaffiltext{4}{Canada
Research Chair}  \altaffiltext{5}{Department of Physics and Astronomy,
McMaster University, 1280 Main St.\ West, Hamilton, Ontario, L8S 4M1,
Canada.}

\begin{abstract} 

One of the most important and poorly-understood issues in structure
formation is the role of outflows driven by active galactic nuclei
(AGN).  Using large-scale cosmological simulations, we compute the
impact of  such outflows on the small-scale distribution of the cosmic
microwave background  (CMB).  Like gravitationally-heated  structures,
AGN outflows induce CMB distortions both through thermal  motions and
peculiar velocities, by processes known as the thermal and kinetic
Sunyaev-Zel'dovich (SZ) effects, respectively.   For AGN outflows the
thermal SZ effect is dominant, doubling the angular power spectrum on
arcminute scales.  But the most distinct imprint of AGN feedback is a
substantial increase in the thermal SZ distortions around elliptical
galaxies, post-starburst ellipticals, and quasars, which is
linearly proportional to the outflow energy.     While point source
subtraction is difficult for  quasars, we show that by
appropriately stacking microwave measurements around early type
galaxies, the new generation of small-scale microwave telescopes will
be able to directly measure AGN feedback at the level important for 
current theoretical models.

\end{abstract}


\section{Introduction}

Large-scale measurements of cosmic microwave background (CMB)
anisotropies have played a central role in the development of the
modern cosmological model.  By giving us a picture of the  universe
during the epoch of linear fluctuations, they have
provided us with a foundation from which to interpret the subsequent
growth of structure.  Thus from the spectacular initial detections
made by the COBE satellite  (Smoot \etal 1992),  to the detailed
flatness constraints provided by balloon-born missions  (Hanany \etal
2000; Mauskopf \etal 2000), to the precision measurements of
cosmological parameters derived from the WMAP  observations (Spergel
\etal  2003; 2007),  our understanding of structure formation has
progressed hand-in-hand with improvements in large-scale CMB
anisotropy  measurements.

On angular scales below $\approx 10',$ the science potential of
CMB  measurements  has yet to be realized.  On these scales, Silk
damping washes out the primary anisotropies, while  a host of
smaller-scale secondary anisotropies are imprinted as the photons
propagate across  the universe.  One of the most important sources of
these effects is the scattering of CMB photons by hot electrons, a
process that was first studied by Sunyaev \& Zel'dovich (1970; 1972).
This process can be divided into two contributions.  The largest of
these is the thermal Sunyaev-Zel'dovich (SZ) effect 
 in which inverse Compton scattering
preferentially increases the  energy of CMB photons passing through
hot and dense regions,  introducing anisotropies with a distinctive
frequency dependence.  A smaller contribution arises from the kinetic
SZ effect in which Doppler scattering in dense regions with
significant peculiar motions induces  fluctuations with the same
frequency dependence as the CMB itself.

The most important source of such hot and dense regions is undoubtedly
the gravitational collapse of gas into large dark
matter halos, and  numerous numerical simulations of the
Sunyaev-Zel'dovich effect from this  process have been  carried out by
several groups (\eg Scaramella, Cen, \& Ostriker 1993;   Hobson \&
Magueijo 1996; da Dilva \etal 2000; Refregier \etal 2000; Seljak,
Burwell, \& Pen 2001; Springel, White, \& Hernquist 2001; Zhang, Pen,
\& Wang 2002; Roncarelli \etal 2007).    Yet dark-matter driven
gravitational heating need not provide the dominant SZ contribution at all
scales and in all environments.  In fact the gas responsible for
the largest SZ distortions, the intracluster medium (ICM) in galaxy
clusters, is observed to have been substantially heated by
nongravitational sources (Cavaliere \etal 1998;  Kravtsov \& Yepes
2000;   Wu \etal 2000; Babul \etal 2002).  In galaxy groups, such
nongravitational effects are even more severe, causing groups with a
large range of X-ray luminosities to all be heated to a similar value
of $ \approx 1$ keV per baryon (\eg Arnaud \& Evrard 1999; Helsdon \&
Ponman 2000).

But perhaps the most dramatic,   yet poorly understood manifestation
of nongravitational heating is in establishing the observed
``downsizing'' (Cowie \etal 1996)  trend in the evolution of  
star-forming galaxies and active galactic nuclei (AGN).  Here the issue is
that since $z \approx 2$ the characteristic mass scale of star-forming
galaxies and the typical luminosities of AGN have dropped by over an
order of magnitude (\eg Arnouts \etal 2005, Treu \etal 2005; Pei 1995;
Ueda \etal 2003; Barger \etal 2005).  This ``antiheirarchical'' trend
is  in direct conflict with  the expectations of the long-standing
model of structure formation  in which gas condensation and heating is
driven purely by dark mater halos, which grow hierarchically by accretion
and merging over time.  In this picture, as gas falls into potential
wells, it is shock heated and must radiate this energy away before
forming stars (Rees \& Ostriker 1977; Silk 1977).  The larger the
structure, the longer it takes to cool, and thus the formation of
history of galaxies is even more hierarchical than the dark matter history.

Although early modeling did not reproduce this downsizing behavior,
the trend has been successfully reproduced in more recent theoretical
models that include  a large source of heating associated with
outflows  from AGN (Scannapieco \& Oh 2004; Binney 2004; Granato \etal
2004; Scannapieco, Silk, \& Bouwens 2005;  Di Matteo, Springel, \&
Hernquist 2005; Croton \etal 2006; Cattaneo \etal 2006; Thacker,
Scannapieco, \& Couchman 2006, hereafter TSC06; Di Matteo \etal 2007).
In this picture, AGN outflows associated with broad-absorption line
winds and radio jets heat the surrounding intergalactic medium (IGM)
to sufficiently high temperatures to prevent it from cooling and thus
from forming further  generations of stars and AGN.  This feedback
requires an energetic outflow, driven by a large AGN, to be effective
in the dense, high-redshift IGM, while in the more tenuous
low-redshift IGM, equivalently long cooling times can be achieved by
less energetic winds.  The lower the redshift, the smaller the  galaxy
that is able to exert efficient feedback, thus resulting in cosmic
downsizing.

However, the details of AGN feedback remain extremely uncertain.
Heating may be impulsive (\eg TSC06), more gradual (\eg  Br\" uggen,
Ruszkowski, \& Hallman 2005; Croton \etal 2006), or modulated
primarily by the properties of the surrounding material (\eg Dekel \&
Birnboim 2006).  Furthermore, direct measurements of the kinetic
energy input from AGN are notoriously difficult  and range from $\sim
1 \%$ or less (de Kool 2001) to $\sim 60\%$  of the AGN's total
bolometric energy (Chartas \etal 2007).  Finally,  several alternative
ideas have been suggested to explain downsizing outside of the context
of AGN feedback altogether (\eg Kere\v{s}, \etal 2005; Khochhfar \&
Ostriker 2007; Birnboim, Dekel, \& Neistein 2007).

Here we show that the definitive measurement resolving this issue may
again come from the microwave background, through observations of SZ
distortions.   As it directly probes the spatial distribution of
heated gas, SZ detections are able to place constraints on the primary
theoretical uncertainty in AGN feedback models, namely the nature and
degree of nongravitational gas heating.  Fortunately, the rise in
importance of this issue for theory is paralleled by
recent advances in observation.  Motivated primarily by deriving
cosmological constraints though the detection of galaxy clusters (\eg
Holder \etal 2000; Majumdar \& Mohr 2003; Battye \& Weller 2003;
Schulz \& White 2003), a large number of blank-field small-scale CMB
surveys are now underway, using telescopes that will push into the
interesting regime for AGN feedback.  These include surveys with the
Atacama Cosmology Telescope  (Kosowsky \etal 2006),  the South Pole
Telescope (Ruhl \etal 2004), the Atacama Pathfinder Experiment (Dobbs
\etal 2006), and  the Sunyaev-Zel'dovich Array (Loh \etal 2005), many
of which will be coordinated to overlap with optical surveys.  These
advances are particularly important as less sensitive  small-scale CMB
surveys have already hinted at an excess of SZ  distortions (Mason
\etal 2003;  Bond 2005; Dawson \etal 2006; Kuo \etal 2007).

In this work we make use of the first large-scale cosmological
hydrodynamical simulation to explicitly include AGN feedback (TSC06)
to construct  detailed maps of their imprint on  the CMB. Analyzing these
maps and comparing them with other observables, we are able to
determine the most promising approaches to using these ongoing surveys
to constrain AGN feedback. Previous studies of nongravitational
heating on the SZ background focused on the impact of starburst
winds at high redshift (Majumdar, Nath, \& Chiba 2001; White,
Hernquist, \& Springel 2002) and signatures from Population III stars
(Oh,  Cooray,\& Kamionkowski 2003).  Recently, Chatterjee \& Kosowsky (2007)
made analytic estimates  of the impact of quasar winds on the CMB
power spectrum and the  cross-correlation of the SZ effect with
optical sources.  Here we show directly from simulations  that while
basic quantifiers such as the power spectrum are difficult to
interpret, approaches based on the cross-correlation of SZ distortions
with optical observations  will soon provide a clean and direct
probe of AGN feedback.

The structure of this work is as follows. In \S 2 we describe our
numerical simulations and the methods used to construct maps of the
CMB distortions from the kinetic and thermal SZ effects.  In \S 3 we
compute the contribution of AGN outflows to the CMB power spectrum,
calculate their impact on the regions surrounding individual galaxies
and AGN, and quantify the sensitivity of  CMB-galaxy cross
correlations to AGN feedback.   Finally, our conclusions are
summarized in \S 4.

\section{Method}

\subsection{Simulations}

In order to disentangle the SZ signatures of AGN feedback  from
distortions due purely to gravitational heating, we  made use of two
simulations, the large  AGN feedback simulation first presented in
TSC06 (using $2\times  640^3$ particles) and a smaller-scale
comparison simulation (using $2\times 320^3$ particles) in which
structure formation,  gas cooling,  and star formation were tracked
exactly as in the fiducial run but no outflows were added.  In both
cases, based on  wide range of cosmological constraints (\eg  Spergel
et al. 2003; Vianna \& Liddle 1996; Riess \etal 1998; Perlmutter \etal
1999), we adopted Cold Dark Matter cosmological model with parameters
$h=0.7$, $\Omega_0$ = 0.3, $\Omega_\Lambda$ = 0.7, $\Omega_b = 0.046$,
$\sigma_8 = 0.9$, and $n=1$, where $h$ is the Hubble constant in units
of 100 km s$^{-1}$ Mpc$^{-1}$, $\Omega_0$, $\Omega_\Lambda$, and
$\Omega_b$ are the total matter, vacuum, and baryonic densities in
units of the critical density, $\sigma_8^2$ is the variance of linear
fluctuations on the $8 h^{-1}{\rm Mpc}$ scale, and $n$ is the ``tilt''
of the primordial power spectrum. Note that these parameters are
slightly different than those preferred by the more recent WMAP data
(Spergel \etal 2007), which was released after our large AGN feedback
run had already been completed.  Initial conditions were computed
using  the Eisenstein \& Hu (1999) transfer function.

In both runs, as in our previous work  (\eg Scannapieco, Thacker, \&
Davis 2001),  simulations were conducted with a parallel OpenMP based
implementation of the ``HYDRA'' code (Thacker \& Couchman 2006)  that
uses the Adaptive Particle-Particle, Particle-Mesh algorithm to
calculate gravitational forces (Couchman 1991), and the smooth
particle hydrodynamic (SPH) method to
calculate gas forces (Lucy 1977; Gingold \& Monaghan 1977).  As the
details of this code and our outflow implementation are described in
detail elsewhere (Scannapieco, Thacker, \& Davis 2001; TSC06) here we
only summarize the aspects most relevant to the SZ effect.

Our study is targeted to relatively large and late forming
structures,  and for both runs we have  kept the metallicity constant at
$Z=0.05$, to mimic a moderate level of enrichment.  Similarly, because
the epoch of  reionization is poorly known and because reionization
has little impact on  mass scales greater than $10^{9}$\msol
(Barkana, \& Loeb 1999), we did  not include a  photoionization
background in either simulation.  In our AGN feedback run, which was
designed to cover a large range in the AGN luminosity function, we used a
simulation box of size $146 h^{-1}$ Mpc filled with $2\times
640^3$ particles, which corresponded to a  dark-matter particle mass of $1.9
\times 10^{8} \msun$ and a gas particle mass of $2.7 \times 10^{7} \msun.$
Our comparison simulation, which was carried out purely for the purposes
of the current study, used $320^3$ particles in a box
$73  h^{-1}$ comoving Mpc on a side,  corresponding to the same particle 
masses as in the AGN run. Both runs were terminated 
at $z = 1.2$, at
which point integration was becoming  expensive due to the
single-stepping nature of the Hydra code, and  impractical in a shared
queue environment.  This means that our results do not contain the SZ
contribution from lower redshifts, during which most galaxy clusters
are formed.    However, as $z=1.2$ is  well past the peak epoch of AGN
activity (eg.\ Ueda \etal 2003;  Barger \etal 2005), our results
should do well  at quantifying the AGN contribution to the SZ effect,
which is our main focus here.

As in TSC06, bright quasar-phase AGN are associated with  galaxy
mergers, which are tracked by labeling gas  particles and  identifying
new groups in which at least 30\% of the accreted mass  does not come
from a single massive progenitor.  Once a merger has been identified,
we compute the mass of the  associated supermassive black hole,
$M_{\rm BH},$ from the circular velocity of the remnant, $v_{\rm c}$
using the observed $M_{\rm BH}- v_{\rm c}$ relation (Merrit \&
Ferrarese 2001; Tremaine \etal 2001; Ferrarese 2002) which gives 
\be
M_{\rm BH} = 2.8 \times 10^8  \left( \frac{v_{\rm c}}{300 \, {\rm km}
\,{\rm s^{-1}}} \right)^5.  \ee Here $v_{\rm c}$ is estimated as  \be
v_{\rm c} = \left[\frac{4\pi}{3} G \rho_{\rm v}(z) r_{\rm v}^2
\right]^{1/2}, 
\ee 
where $G$ is the gravitational constant,
$\rho_{\rm v}(z)$ is the virial density as a function of redshift, and
$r_{\rm v}$ is the implied virial radius for a group of $N$ gas
particles with mass  $m_{\rm g}$ 
\be r_{\rm v}=\left[ {N  m_{\rm g}
\Omega_0/\Omega_b \over 4/3 \pi \rho_{\rm v}(z)}  \right]^{1/3}.  
\ee
Following Wyithe \& Loeb (2002) and Scannapieco \& Oh (2004), we assume
that for each merger the accreting black hole shines at its Eddington
luminosity ($1.2 \times 10^{38}$ ergs s$^{-1}$ $\msun^{-1}$) for a time
taken to be a fixed fraction of the dynamical time of the system,
$t_{\rm d} = 0.055 r_{\rm v}/v_{\rm c} =
 5.8 \times 10^{-3} \Omega(z)^{-1/2} H(z)^{-1}$.  In TSC06 it was
demonstrated that, apart from a discrepancy for the very luminous end,
this simple model does extremely well at reproducing the observed AGN
luminosity function as well as the large and small-scale clustering of AGN
over the full range of simulated redshifts. Furthermore, the discrepancy
at the very  luminous end for the simulation as compared to the semi-analytic 
model  could be attributed to the relative efficiency of
shock heating on  substructure (e.g. see Agertz \etal 2006 for a
discussion of numerical  "stripping" issues). 
Note that while our approach does not distinguish between AGN formed
in gas-rich ``wet'' mergers and gas-poor ``dry'' mergers, unlike
at lower redshift (\eg Bell \etal 2006), dry mergers are likely to
be relatively unimportant at the $z \ge 1.2$ redshifts we are studying.

In our fiducial run, we also assume that a fixed fraction
$\epsilon_{\rm k} =0.05$ of the bolometric energy of each AGN is
channeled into a kinetic outflow.   This value is consistent with
other literature estimates  (\eg Furlanetto \& Loeb 2001; Nath \&
Roychowdhury 2002), as well as observations (Chartas \etal 2007).
Each outflow is then launched with an energy input of  
\be 
E_{\rm k}=6
\times10^{36}\; \left({ M_{\rm bh} \over  \msol}\right) 
\left({t_{\rm d} \over  {\rm s}}\right) \;{\rm ergs}.   
\label{eq:ekin}
\ee  
Given the uncertainties
surrounding AGN outflows, we simply model each expanding outflow as a
spherical shell at a radius $2r_{\rm v}$  which is constructed by
rearranging all the gas within this radius, but outside $r_{\rm v},$
and  below a density threshold of $2.5\rho_{\rm v}.$ Finally, the
radial velocity $v_s$ and temperature $T_s$  of the shell are
determined by fixing the postshock temperature, $T_s$, to be $T_s = 13.6
K [v_s/({\rm km} {\rm s}^{-1})]^2$ and choosing $v_s$ such that the
sum of the thermal and kinetic energy of the shell equals $E_{\rm k}$
minus the energy used to move particles from their initial positions
into the  shell.  As shown in TSC06, this prescription results in
a level of preheating in galaxy clusters and groups that is in good
agreement with observations.

\begin{figure}
\centerline{\psfig{figure=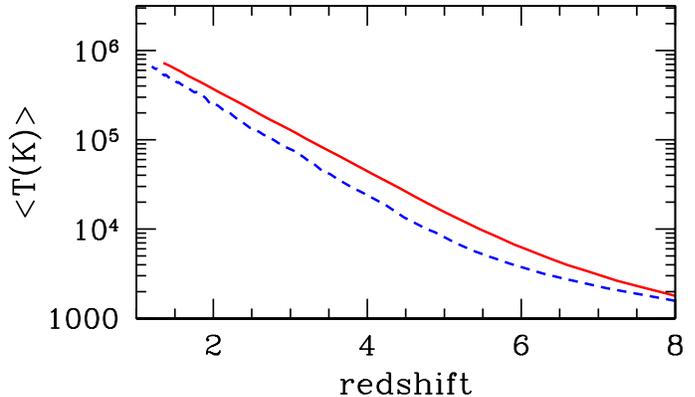,height=5.8cm}}
\caption{Evolution of the mass-averaged gas temperature in our AGN feedback
(solid) and comparison (dashed) runs.\\}
\label{fig:norm}
\end{figure}

In Figure \ref{fig:norm}, we show the mass-averaged gas temperature in our 
AGN feedback and  comparison runs.  The presence of outflows has a
substantial impact on the IGM, raising the mean gas temperature in the
simulation by  $\approx 50 \%$.  Furthermore, as discussed in TSC06,
most of this heating is targeted  toward  the densest regions,
exactly those that contribute most to the SZ effect.\\ \\

\subsection{Construction of SZ maps}

As the thermal and kinetic SZ effect have different frequency
distributions, we calculate each of them separately from our
simulations.  In the (nonrelativistic) thermal SZ case,  the change in
the temperature of the CMB as a function of frequency is  given by 
\be
\frac{\Delta T}{T} = y \left[ x\frac{e^x+1}{e^x-1} -4 \right],
\label{eq:DT}
\ee where $x = h \nu/k T_{\rm CMB}$ is the dimensionless frequency,
and the Compton $y$ parameter is defined as 
\be 
y \equiv \sigma_T \int
dl   \, n_e  \frac{k(T_e-T_{\rm CMB})}{m_e c^2},
\label{eq:y}
\ee  
where $\sigma_T$ is the Thompson cross section,  $m_e$ is the mass
of the electron,  $n_e$ is the density of electrons, $T_e$ is the
electron temperature,  $T_{\rm CMB}$ is again the temperature  of the
CMB, and the integral is performed over the proper distance along the
line of sight.  Note that $y$ is simply a rescaled version of the
line-of-sight integral of the pressure.  Note also that
in the Rayleigh-Jeans limit,  
in which $x \ll 1,$ eq.\ (\ref{eq:DT}) reduces to $-2 y,$
and we focus on this limit throughout our discussion below.

For the kinetic SZ effect, on the other hand, the magnitude of the CMB
distortions is frequency independent and $\Delta T/T = - b$. In this
case 
\be 
b \equiv \sigma_T \int dl \, n_e \frac{v_r}{c},
\label{eq:b}
\ee where $v_r$ is the radial peculiar velocity of the gas, and the
integral is again performed along the line of sight.  

To construct maps of these distortions from
our simulations, we followed a method similar to da Silva \etal
(2000) and Springel \etal (2002).  For each simulation output, we smoothed
the density, temperature, and velocity fields onto a cubic
grid.  The  smoothing procedure used the standard SPH smoothing
methodology, that cells at a distance  $|{\bf r}_{cell}-{\bf r}_i|$
from a particle $i$, are incremented by 
\begin{equation}
A_{cell} \equiv {1 \over w_{n_i}} { A_i \over \rho_i} m_i W(|{\bf
r}_{cell}-{\bf r}_i|,h_i),
\end{equation}
where $A_i$ is the scalar field value for particle $i$, $\rho_i$ is
the density of particle $i$, $W$ is the B2-spline kernel, and $w_{n_i}$
is a  normalization factor. The smoothing process generalizes in the
natural  fashion  for vector fields. The normalization factor is
required to  ensure  that the weighting of the assigned kernel is
correct when summed  across a finite number of cells (the standard
kernel definition  normalizes to unity over a continuous spherical
volume of radius  $2h_i$). If a  particle is smoothed  over $n_{cell}$
cells within a spherical volume specified by the  radius $2h_i$,  then
$w_{n_i}$ is given by
\begin{equation}
w_{n_i} = \sum_{k=1}^{n_{cell}} W(|{\bf r}_{cell_k}-{\bf r}_i|,h_i),
\end{equation}
where we have used $k$ to distinguish that the summation is over
cells,  rather than over particles.

Once the smoothed grid had been constructed it was  then projected in
the $x$, $y$ and $z$ directions. Since projection amounts to a
summation along each axis  direction the smoothing process could
potentially have been performed in  two dimensions alone. However, we
project the three dimensional grid to ensure  that the integration in the
projection axis accounted for the quantization effects associated with
smoothing onto a fixed number of  cells.
Next we  constructed a grid of spanning ($1.1$ deg)$^2$
and made up of  1024$^2$ rays in the fiducial run, and spanning ($0.55$
deg)$^2$ and made up of 512$^2$ rays in the comparison run, such that in
both  cases the full simulation subtended an angle equal to the
field at our highest redshift output of $z=10.$  Finally we projected
along each of our sightlines, choosing a random translation and
orientation at every redshift, and weighting the slices according to
eqs.\ (\ref{eq:y}) and (\ref{eq:b}).

\section{Results}

\subsection{Overall Properties}

\begin{figure}
\centerline{\psfig{figure=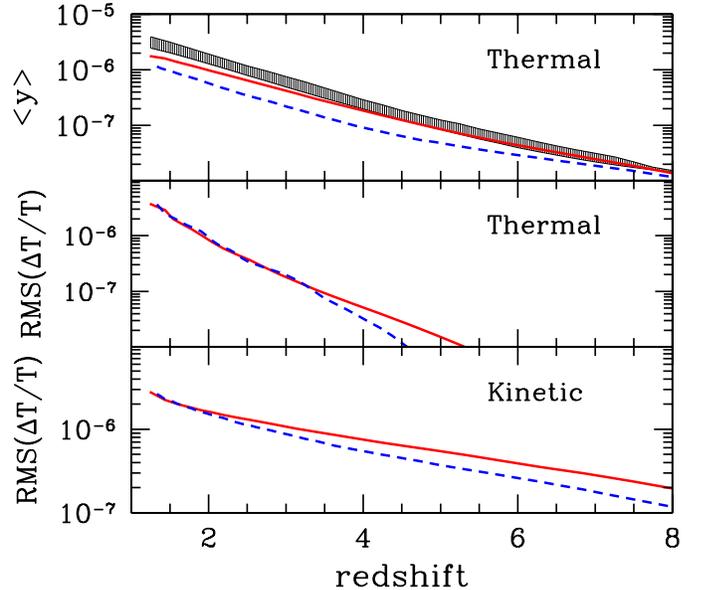,height=8.5cm}}
\caption{{\em Top:} Total integrated value of the  evolution of the
average Compton $y$ parameter as a function of redshift for the AGN
feedback run (solid line) and comparison run (dashed line) The
shaded region is bounded below by the Scannapieco \& Oh (2004)
predictions for the $y$ distortion due to AGN feedback (with
$\epsilon_k = 0.05$) and from above by the sum of this prediction plus
the results of our comparison run.  {\em Middle:} 
Total RMS value of $\Delta T/T = -
2y$ from the thermal SZ effect, as a function of redshift, with lines
as in the upper panel.  {\em Bottom:} 
Total RMS $\Delta T/T$ from the kinetic SZ effect as a function of redshift.\\}
\label{fig:evolution}
\end{figure}

In Figure \ref{fig:evolution}, we study which redshifts contribute
most to CMB distortions, by integrating from $z=10$ (our assumed
redshift  of reionization) down to various redshifts.   From eq.\
(\ref{eq:y}) the change in the average Compton $y$ distortion per unit
$dl$ is proportional  to the mass-average temperature. Thus,
consistent with the temperature evolution shown in Figure
\ref{fig:norm},  $\left< y \right>$ is roughly $50 \%$ higher in the
AGN feedback run at all redshifts.  Furthermore, for most redshifts,
this increase is similar to the one derived from analytic estimates as
presented in Scannapieco \& Oh (2004). In fact, the difference between this
model and our simulations may be
due to the presence of small-scale structure around AGN outflows,
which were ignored in the analytic estimates, and whose mixing with
outflow material may be somewhat under-resolved in the simulations (see
TSC06 for a full discussion of this issue).

Yet, the average SZ distortion is not the most easily detectable
quantity.  Rather, the majority of CMB measurements are differential
in nature, and sensitive to small changes on top of an overall
background that is much less well measured.  To quantify these
differences, in the center and low panels of Figure 2, we plot the RMS
scatter in $\Delta T/T$ due to the thermal and kinetic SZ effects
respectively.  In both cases the changes in the amplitude of
the distortions are small, indicating that AGN have a minor impact on
the overall variance of the temperature and velocity.

\begin{figure}
\centerline{\psfig{figure=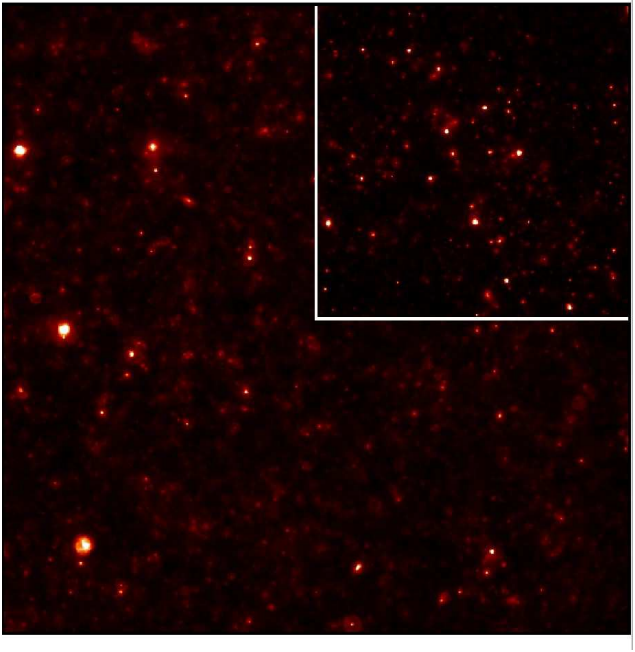,height=9.0cm}} %SZmap.eps
\caption{Map of the thermal SZ effect in our AGN feedback and
comparison runs (large panel and inset panel, respectively).  Here the
area of the largest panel is $(1.1 {\rm deg})^2$ and the color scale
runs from $\frac{\Delta T}{T} = 0$ (black) to $-5 \times 10^{-5}$
(white), (or 0 to -135 $\mu$K).\\}
\label{fig:SZmap}
\end{figure}

\begin{figure}
\centerline{\psfig{figure=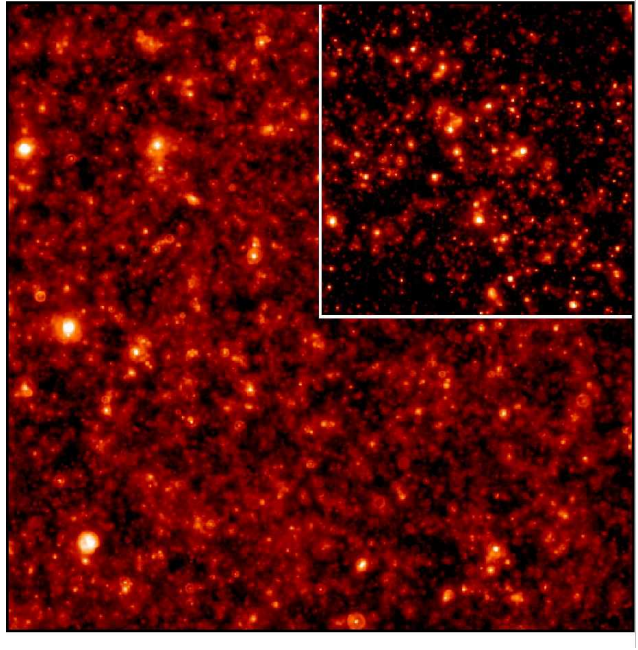,height=9.0cm}} %SZlogmap.eps
\caption{Map of the logarithm of the thermal SZ effect, ranging from
 $\log(-\Delta T/T) = \log(1 \times 10^{-6})$ (black) to $\log(-\Delta
 T/T) = \log(5 \times 10^{-5})$ (white), with panels as in Figure
 \protect\ref{fig:SZmap}.  While the AGN simulation has a higher
 average Compton distortion, structure has been smoothed  out on the
 smallest scales.\\}
\label{fig:SZmap2}
\end{figure}

In Figure \ref{fig:SZmap} we directly compare maps of the thermal SZ
distortions from  both these runs, integrated down to our final
simulation redshift of $z=1.2.$ The maps clearly indicate a
lack of small hot regions in the AGN feedback run relative to the
comparison run. To help bring out the structure in these maps, we also
plot the same data on a logarithmic scale in Figure \ref{fig:SZmap2}.
Here we see that the AGN feedback  run has a higher average level of
distortions, consistent with the $\left<y\right>$ evolution  in Figure
\ref{fig:evolution}.  Furthermore, while it is difficult to tell the
overall level of the variance between the maps, it is clear that the
scale of the structures is quite different.  In particular, while the
AGN feedback simulation has fewer small pockets of hot gas, these are
compensated for by a number of larger and more diffuse heated regions.

\begin{figure}
\centerline{\psfig{figure=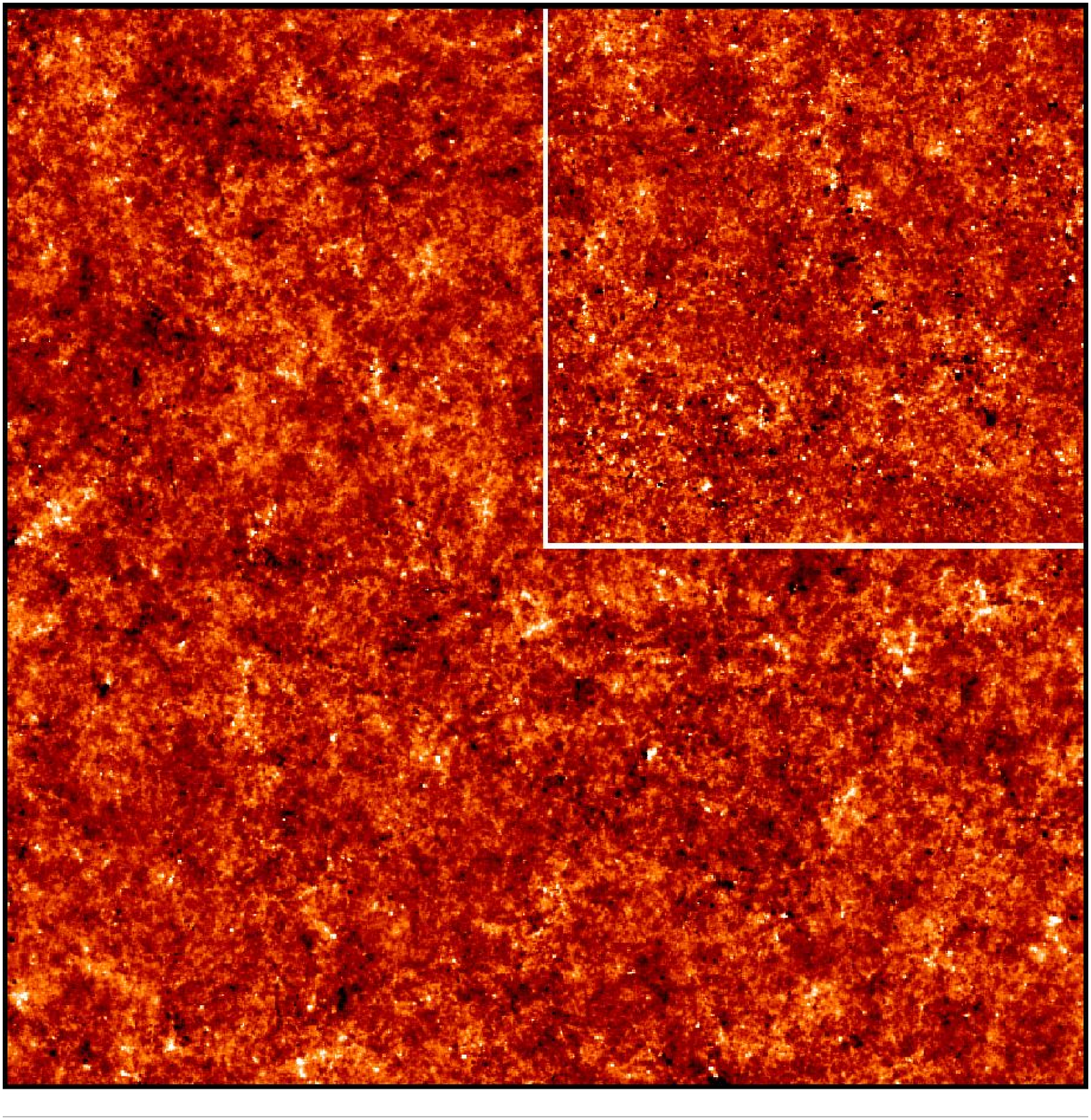,height=9.0cm}} %kSZmap.eps
\caption{Map of the kinetic SZ effect ranging from  $\frac{\Delta
T}{T} = -1 \times 10^{-5}$ (black, moving away from the observer)  to
$1 \times 10^{-5}$ (white, moving towards the observer), that is -27.3
to 27.3 $\mu$K.\\}
\label{fig:kSZmap}
\end{figure}

Finally, the kinetic SZ maps, as shown in Figure \ref{fig:kSZmap}, look
similar  in the two simulations.   The differences in this
case appear to be quite subtle, and must be distinguished by a more
detailed, quantitative approach. \\ \\ \\ \\

\subsection{Power Spectrum}
\subsubsection{Results from Simulated Maps}
The most common measure of CMB fluctuations is  the angular power
spectrum, obtained by decomposing the temperature distribution on the
sky into spherical harmonics, $\Delta T/T({\bf \hat n}) =
\sum_{\ell,m} a_{\ell,m} Y_{\ell,m}({\bf \hat n}),$  and carrying out
an average over the coefficients, $C_\ell = (2\ell +1)^{-1}
\sum_{m=-\ell}^{\ell} a_{\ell,m} a^*_{\ell,m}.$  For a small field of
view, as is the case for our constructed map, this is equivalent to
performing a Fourier Transform to obtain \be  {\tilde {\frac{\Delta
T}{T}}} ({\vec \kappa}) = \int d^2 {\vec \theta} \exp \left(-i {\vec
\kappa} \cdot {\vec \theta} \right) {\frac{\Delta T}{T}} ({\vec
\theta}), \ee (where $\vec \theta$  and $\vec \kappa$ are two
dimensional vectors in the plane of the sky) and then carrying out an
azimuthal average to obtain  \be C_\ell = P_{\rm ang}(\kappa = \ell) =
\frac{1}{2 \pi} \int d \phi  {\tilde {\frac{\Delta T}{T}}}
(\kappa,\phi)  {\tilde {\frac{\Delta T^*}{T}}} (\kappa,\phi),  \ee
where ${\vec \kappa} = (\kappa \cos \phi ,\kappa \sin \phi).$ Finally,
as an additional check one can use the fact that the mean  squared
temperature variance is equal to $\sum_\ell (2 \ell+1)  C_\ell/(4
\pi).$

\begin{figure}
\centerline{\psfig{figure=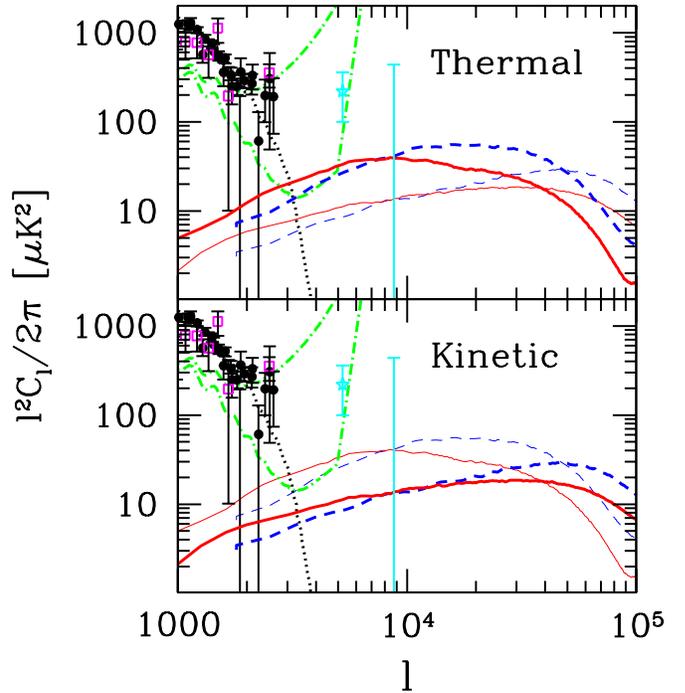,height=10.0cm}}
\caption{{\em Top:} Power spectrum of the thermal SZ distortions in
the AGN feedback run (thick solid line) and comparison run (thick
dashed line), integrated down to $z=1.2$.   The open squares are
measurements by the CBI experiment (Bond \etal 2005),the solid
points are measurements by the ACBAR experiment (Kuo \etal 2007),
and the stars are measurements from 
the BIMA CMB Anisotropy Survey (Dawson \etal 2006).
These are plotted around the primary anisotropy signal (dotted lines). 
For comparison, the thin
solid and dashed lines show the kinetic SZ anisotropies for the AGN
feedback and comparison run, as in the bottom panel.  The dot-dashed
lines show the level of anisotropies detectable by an experiment
spanning a (20 deg)$^2$ area with a $\theta_{\rm FWHM} = 2$ arcmin beam,
with a noise level of (10 $\mu$K)$^2$ per arcmin$^2$ (upper line) and
with a $\theta_{\rm FWHM} = 1$ arcmin with a noise level of (2
$\mu$K)$^2$ per arcmin$^2$ (lower line).  {\em Bottom:} Power spectrum
from the kinetic  SZ distortions in  the AGN feedback run (thick solid
line) and comparison run (thick dashed line), integrated down to
$z=1.2$.  Points and dot-dashed lines are as in the upper panel, and
the thin lines are the thermal SZ  distortions, shown for comparison.\\}
\label{fig:cl}
\end{figure}

In Figure \ref{fig:cl}  we plot the angular power spectrum  from our
AGN feedback and comparison simulations, averaging over 16
realizations of the maps:  (4.4 deg)$^2$ in the AGN feedback case and
(2.2 deg)$^2$ in the comparison case.  As expected from a visual
inspection of Figures \ref{fig:SZmap} and \ref{fig:SZmap2} , AGN
feedback has a clear impact on the thermal SZ signal, smoothing the
structures at the smallest scales ($\ell \ge 10^4$) while at the same
time adding large heated regions that increase the overall
fluctuations at larger scales ($\ell \le 10^3$).  On the other hand,
AGN feedback has only a weak impact  on the kinetic SZ effect,
slightly smoothing the smallest structures, without having a
noticeable impact on larger scales.  To some extent,  the lack of this
large-scale signal may be due to our  choice of initially spherical
outflows, which should not significantly add to the total integrated
column of material moving towards or away from the observer along any
line of sight.  If outflows were extremely asymmetric, however, from
eqs.\ (\ref{eq:y}) and (\ref{eq:b}) the ratio of the distortions in
the kinetic  to the thermal SZ effects would go as \be  \frac{b}{y}
\approx \frac{v/c}{{kT}/{m_e c^2}} \approx \frac{m_e}{m_p (v/c)}.
\ee  For a  typical outflow velocity of $ v \approx 1000$ km/s, this
gives  $b/y \approx 1/6$, such that the even in this extreme case,
the increase in $C_\ell$s due to the kinetic effect would be at least
$\approx 30$ times smaller than the thermal signal.

However, even in the thermal case, the increase in the large-scale
angular  power spectrum is not dramatic.  For comparison, in Figure
\ref{fig:cl}   we have also plotted the primary CMB signal, as well as
a  summary of recent  measurements from the CBI and ACBAR
experiments.  From these points it is clear that the secondary
anisotropies seen in current experiments are not  likely to have been
imprinted by SZ distortions at $z \ge 1.2$, nor does the possible
excess of small-scale power seen in the CBI experiment (see however,
Kuo \etal 2007) appear to be related to AGN feedback.

\subsubsection{Required Sensitivity and Resolution to detect AGN 
distortions}

Pushing to the near future, however, the increase in the angular power
spectrum due to AGN feedback is well within the range of upcoming
experiments.  To calculate these sensitivities we make use  the simple
estimate from  Jungman \etal (1996), which gives the RMS noise at
multipole $\ell$ as 
\be 
\sigma_\ell  = \left[ \frac{2}{(2 \ell + 1)
f_{\rm sky}} \right]^{-1/2}  \left[C_\ell + (w f_{\rm sky})^{-1}
e^{\ell^2 \sigma_b^2} \right], 
\ee 
where $w \equiv (\sigma_{\rm pix} \theta_{\rm FWHM})^{-2}$ is the
 weight per solid angle,  a
pixel-size independent measure of the noise that is calculated from
the full-width at half-max of the Gaussian beam, $\theta_{\rm FWHM},$
and the noise variance per $\theta_{\rm FWHM} \times \theta_{\rm FMWH}$ pixel,
$\sigma_{\rm pix}^2.$  Finally,  $\sigma_b = 7.42 \times 10^{-3} \times
\theta_{\rm FHWM}$ is a measure of the beam size and $f_{\rm sky}$
is the fraction of the sky covered.

Comparing these measurements with the $C_\ell$s from our simulations,
we see that the excess due to AGN feedback is easily measurable
by an experiment with a typical full-width at half max
beam size of  $\theta_{\rm FWHM} \approx 1$ arcmin and a noise level
of about (2 $\mu$K)$^2$ per arcmin$^2,$ scanning over a (20 deg)$^2$
patch of the sky.  However, the most convincing constraints on AGN
feedback are not likely to come from statistical constraints derived
purely from the microwave background.

\subsection{Cross Correlations}

Unlike the primary microwave background signal, the SZ distributions are not
well described by a Gaussian random field.  Rather  they contain a wealth
of information beyond the angular power spectrum.  In fact, as is
particularly clear from Figures \ref{fig:SZmap} and \ref{fig:SZmap2},
the thermal SZ distribution is much  more reminiscent of
lower-redshift galaxy surveys than $z \approx 1000$ measurements.
Furthermore, the structures seen in these maps are imprinted
by galaxy activity, meaning that they are strongly
correlated with the positions of sources in large-field optical and
infrared surveys.  With this in mind, we considered the
cross-correlation of our maps with three types of optical sources:
\begin{itemize}
\item quasars, identified as newly-formed black holes that shine 
for
$t_d = 0.055 r_v/v_c$ of the dynamical time of the system (as in
\S2.1)
\item post-starburst elliptical (E+A) galaxies, identified as
mergers observed within 200 Myrs of coalescence
\item  elliptical
galaxies, identified as  mergers observed at any time after coalescence.
\end{itemize}

\begin{figure*}
\centerline{\psfig{figure=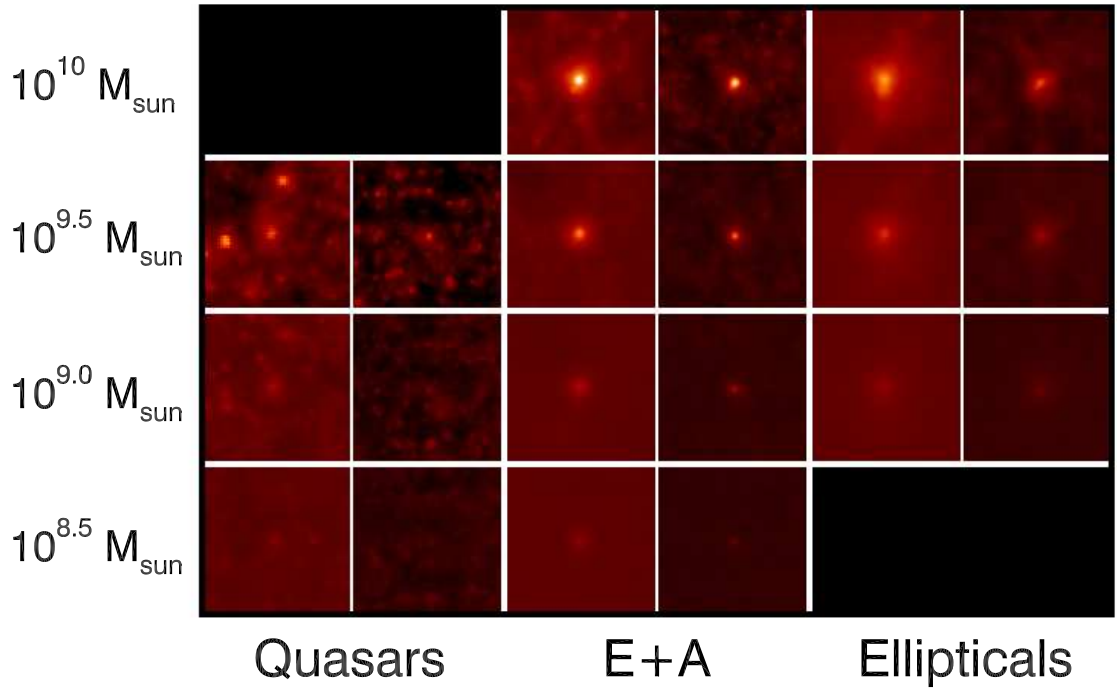,height=10.0cm}} %postage.eps
\caption{Postage stamp images of the thermal SZ effect in a $6 \times
6$ arcmin$^2$ region around  various objects.  Each pair of panels
shows the distortions in the AGN run (left) and comparison run (right)
averaged over all the sources in a (2.2 deg)$^2$ region of the sky.
From top to bottom, the rows are labeled by the black hole masses,
which range from $10^{10} \msun$ to  $3 \times 10^8 \msun$.  From left
to right each of the pairs correspond to  quasars,
post-starburst ellipticals, and all elliptical galaxies. All panels
are shown  using a log scale from $\log(-\Delta T/T) =  \log(2 \times
10^{-5})$ (white) to $\log(-\Delta T/T) = \log(2 \times 10^{-6})$
(black).\\}
\label{fig:postage}
\end{figure*}

In Figure \ref{fig:postage} we show  the results of coadding the
thermal SZ distortions  around each of these objects as selected in
a (2.2 deg)$^2$ region, made up of  4 maps in the AGN feedback run and
16 maps in the comparison run.   The most obvious feature in this plot
is the overall higher level of the SZ background in the AGN feedback
run, a feature that is difficult to observe directly, as
discussed in \S 2.1.  Beyond this overall offset however, these images
also uncover a substantial ``halo''   of SZ distortions, 
which is much higher in amplitude and more spatially 
extended in the AGN feedback case.  While this excess is somewhat
lost in the noise for quasars, which are the rarest sources, it is
clearly and dramatically present in the coadded maps of the much more
numerous E+A and elliptical galaxies. As expected from our discussion above,
constructing similar coadded images of the kSZ effect yielded 
differences that were at least an order of magnitude smaller.

\begin{figure}
\centerline{\psfig{figure=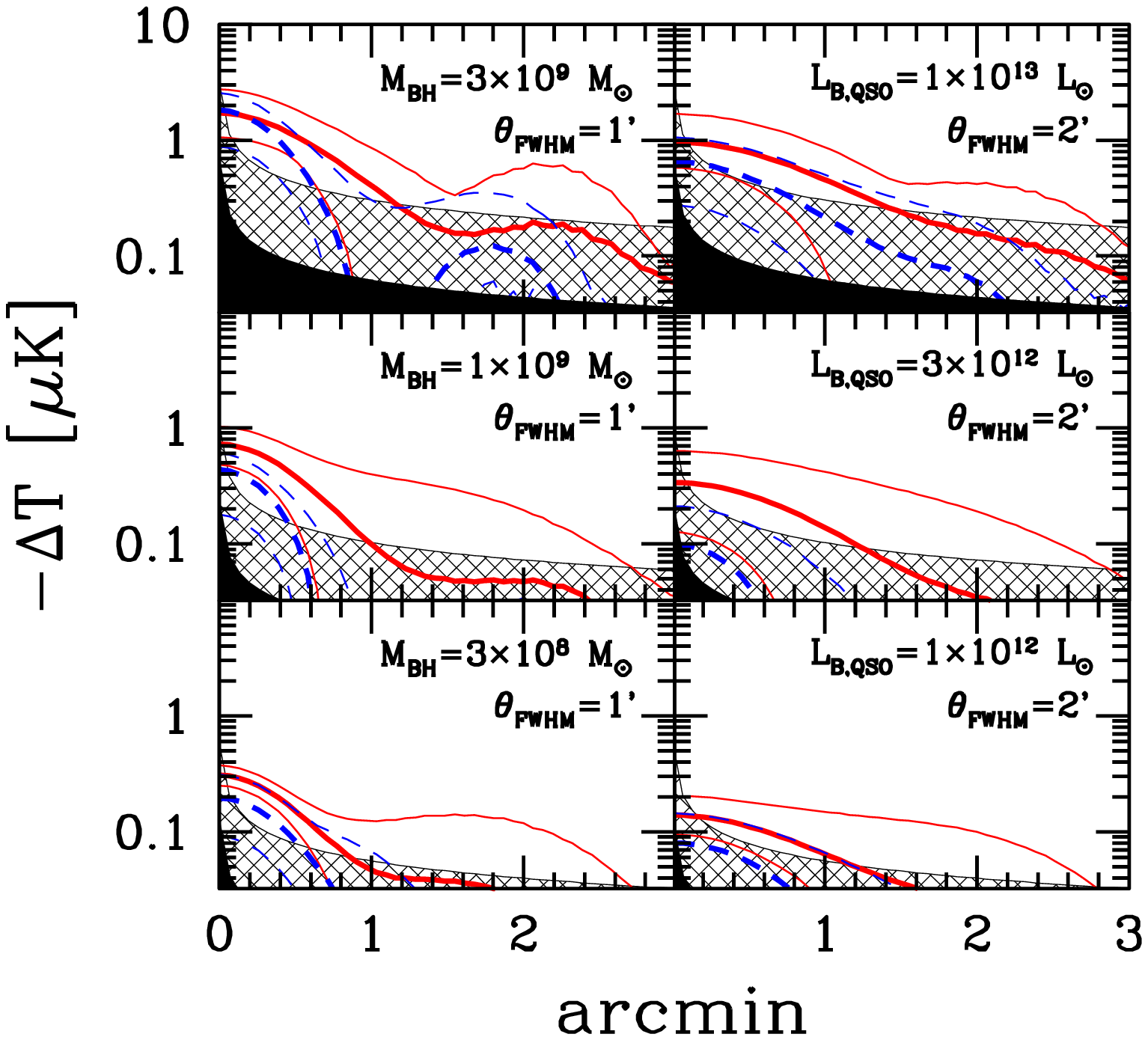,height=9.0cm}}
\caption{Average excess thermal SZ distortions around quasars
in our simulations.  In each panel the thick solid line shows $\Delta
T$ in the AGN  feedback run, averaged over a (2.2 deg)$^2$ area of the
sky.  As a measure of the variance over the sky, the thin solid line shows
the maximum and minimum  $\Delta T$ signal in single (1.1 deg)$^2$
quadrants of this (2.2 deg)$^2$ map.  Similarly, the thick dashed line
shows the average $\Delta T$ in (2.2 deg)$^2$ of sky from the
comparison run, bracketed by the maximum and minimum signals in
quadrants.  Finally, the shaded region shows the noise level
attainable by averaging over all sources in (20 deg)$^2$, for an
experiment with a noise-per-pixel of  (10 $\mu$K)$^2$ per arcmin$^2$
(cross-hatched regions) and (2 $\mu$K)$^2$  per arcmin$^2$ (solid
regions).  The left panels show distortions as observed with a
$\theta_{\rm FWHM} = 1'$ beam and the right panels show the same
distortions for a $\theta_{\rm FWHM} = 2'$ beam. From top to bottom
the panels correspond to black hole masses of $3 \times 10^9 \msun$,
$1 \times 10^9 \msun$, and $3 \times 10^8 \msun,$ respectively,
as marked on the left panels,
and the equivalent B-band luminosity is given in the right  panels.\\}
\label{fig:50}
\end{figure}

To further study the detectability of  the much more promising thermal
signal we processed each of the images in Figure \ref{fig:postage}
in a manner similar to how
one might work with real data sets in the near  future.  First, we
shifted each of them by the overall average $y$ distortion to
account for the difficult  to observe offset between the AGN feedback
and comparison cases.  Next, to approximate the angular resolution of
the next generation of CMB telescopes, we convolved each image with
Gaussian beams with $\theta_{\rm FWHM}$ of 1 and 2 arcmins.  Finally,
as coadding sources washes out any rotational asymmetries, we performed
an azimuthal average to reduce noise while retaining the same
information.

The results of this procedure are shown in Figure \ref{fig:50}, for
quasars with B-band luminosities above  $1\times 10^{13} L_\odot$,  $3
\times 10^{12} L_\odot,$ and $1 \times 10^{12} L_\odot$, which
correspond to   black hole masses of $3 \times 10^9 \msun$, $1 \times
10^9 \msun$, and $3 \times 10^8 \msun.$  Each panel in this plot shows
not only the average signal over all sources in a (2.2 deg)$^2$ region
of the sky, but the maximum and minimum signal in any of the four
($1.1$ deg)$^2$ quadrants in this region, to give an impression of the
variation in this quantity across the sky.  While signals can vary
substantially between quadrants, in all cases the overall average SZ
signal is significantly higher in the AGN feedback case.  
As an estimate of the sensitivity of future experiments
we also include on this plot a background noise level calculated as
\be \sigma_{\rm source}(\theta)  = \left[\frac{\sigma^2_{\rm pix}}
{N_{\rm sources} (2 \pi \theta d\theta)} \right]^{1/2}, \ee where
$d\theta$ is the 1.1 deg/1024 = 0.0644 arcmin pixel size of our maps,
$N_{\rm sources}$ is the number of sources that would be found in a
(20 deg)$^2$ region of sky, and the pixel noise level, $\sigma_{\rm
pix},$ is again estimated as (10 $\mu$K)$^2$ per arcmin$^2$
(cross-hatched region) and (2 $\mu$K)$^2$ per arcmin$^2$  (solid
region).

The impact of AGN outflows is well above the noise for all sources, even if
$\theta_{\rm FWHM} = 2'$ and $\sigma_{\rm pix}^2$ =  (10 $\mu$K)$^2$ per 
arcmin$^2$.   However, in most cases, the profile of the SZ signal above the noise is
indistinguishable from a point source (which would appear on this plot as an inverted
parabola that drops by a factor of two at a distance of $\theta_{\rm FWHM}/2$).
At the same time quasars themselves are often significant 
microwave point sources.   To estimate this intrinsic contribution we convert 
the flux per unit frequency, $F_\nu$, to CMB temperature units as 
\ba
\Delta T &=& \left(\frac{d B_\nu}{d T} \right)^{-1} \
\frac{F_\nu}{\theta_{\rm FWHM}^2}  \nonumber \\
&=& \frac{1 \mu {\rm K}}{0.0084 \, {\rm mJy}} 
\left(\frac{1 \, {\rm arcmin}}{\theta_{\rm FWHM}} \right)^2 F_\nu
\frac{(e^x-1)^2}{x^4 e^x},
%99.27 Jy per steradian is
%8.4e-6 Jansky per armin^2
%and that is about 10-28 ergs/sec/cm^2/Hz
\ea
where $B_\nu$ is the Planck Function and,
as in eq.\ (\ref{eq:DT}), $x \equiv h \nu/ k T_{\rm CMB}$ (Scott \& White 1999).
The ratio of $\nu L_\nu$ at microwave wavelengths to the quasars bolometric luminosity 
is\ $\sim 1/1000$ for ``radio loud'' objects which make up about $10\%$ of the population,
and several orders of magnitude less for radio-quiet objects (Elvis \etal 1994).   
Estimating a typical luminosity distance
to a quasar as $\approx$ 3000 Mpc $h^{-1}$ and considering a typical observing
frequency of 100 GHz this gives
\be
%3000Mpc^2*4*pi*10^11=10^68
\Delta T \approx
%\frac{\rm \nu L_\nu}{10^{40} {\rm ergs \, s^{-1}}}
%\left(\frac{1 \, {\rm arcmin}}{\theta_{\rm FWHM}} \right)^2 \mu {\rm K}, 
%\approx
\frac{M_{\rm BH}}{10^6 \msun}
\left(\frac{1 \, {\rm arcmin}}{\theta_{\rm FWHM}} \right)^2 \mu {\rm K}, 
\ee
which is much larger than the thermal SZ signal.

\begin{figure}
\centerline{\psfig{figure=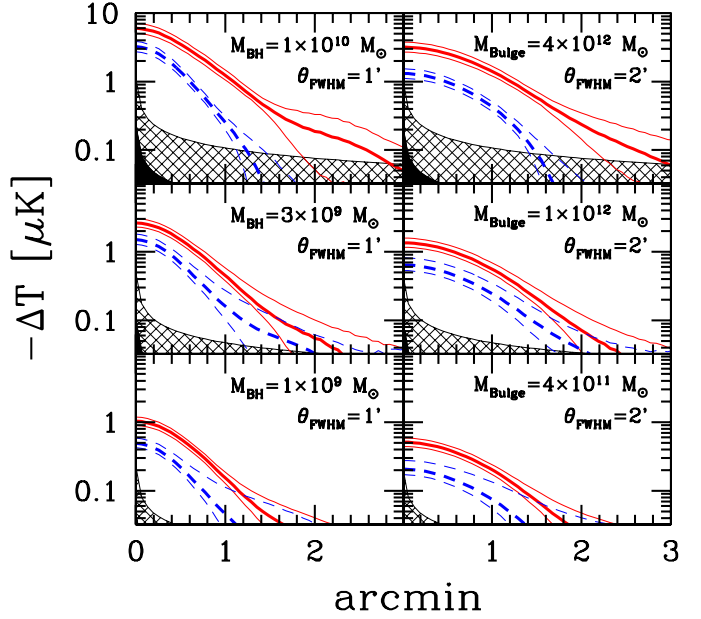,height=9.0cm}}
\caption{Average excess thermal SZ distortions around post-starburst
galaxies in our  simulations, observed within 200 Mys of a merger.  As
there are many more of these objects than quasars, we select slightly
larger objects in this figure than in Figure \ref{fig:50}.  Thus from
top to bottom the panels correspond to  black hole masses of $1 \times
10^{10} \msun$, $3 \times 10^9 \msun$, and $1 \times 10^9 \msun$ 
(marked on the left panels), or equivalent
bulge masses of $4 \times 10^{12} \msun$,  $1 \times 10^{12} \msun,$
and $4 \times 10^{11} \msun$  (marked on the right panels).  
Otherwise the panels,
lines, and shaded regions are defined as in Figure \ref{fig:50}.
Thus, the left panels show distortions as observed with a $\theta_{\rm
FWHM} = 1'$ beam and the right panels show the same distortions for
a $\theta_{\rm FWHM} = 2'$ beam; The thick solid (AGN feedback) and
dashed (comparison) lines show the average value of the SZ distortions
averaged over all sources in a (2.2 deg)$^2$ region of the sky,
bracketed by the maximum and minimum signal in (1.1 deg)$^2$
quadrants; and the shaded regions show the noise levels attainable by
averaging over (20 deg)$^2,$ with instrument noise as in Figure
\ref{fig:50}.\\}
\label{fig:200}
\end{figure}

While this can be reduced substantially by selecting radio-quiet
quasars, a better approach is to work with objects observed at later
times after the merger.   In Figure \ref{fig:200} we show the
azimuthally averaged excess thermal SZ contribution around
post-starburst galaxies, identified as mergers observed after the
quasar  phase, but within 200 Myrs after coalescence.   Here we have
converted each black hole mass to its corresponding stellar bulge mass
using a factor of $\approx 400$ as obtained from the analysis in
Marconi \& Hunt (2003). Working with these sources has two main
advantages.  Firstly, as they are much more common, one can consider
slightly larger bulges and hence larger and more spatially extended SZ
distortions,  while at the same time improving the number of sources
that can be coadded in a fixed region of the sky.  Furthermore, as the
nuclear luminosities are typically more than 100 times less during
this phase (\eg Hopkins \etal 2007), point source subtraction is much
easier, although some exclusion of radio-loud sources may still be
necessary.

\begin{figure}

\centerline{\psfig{figure=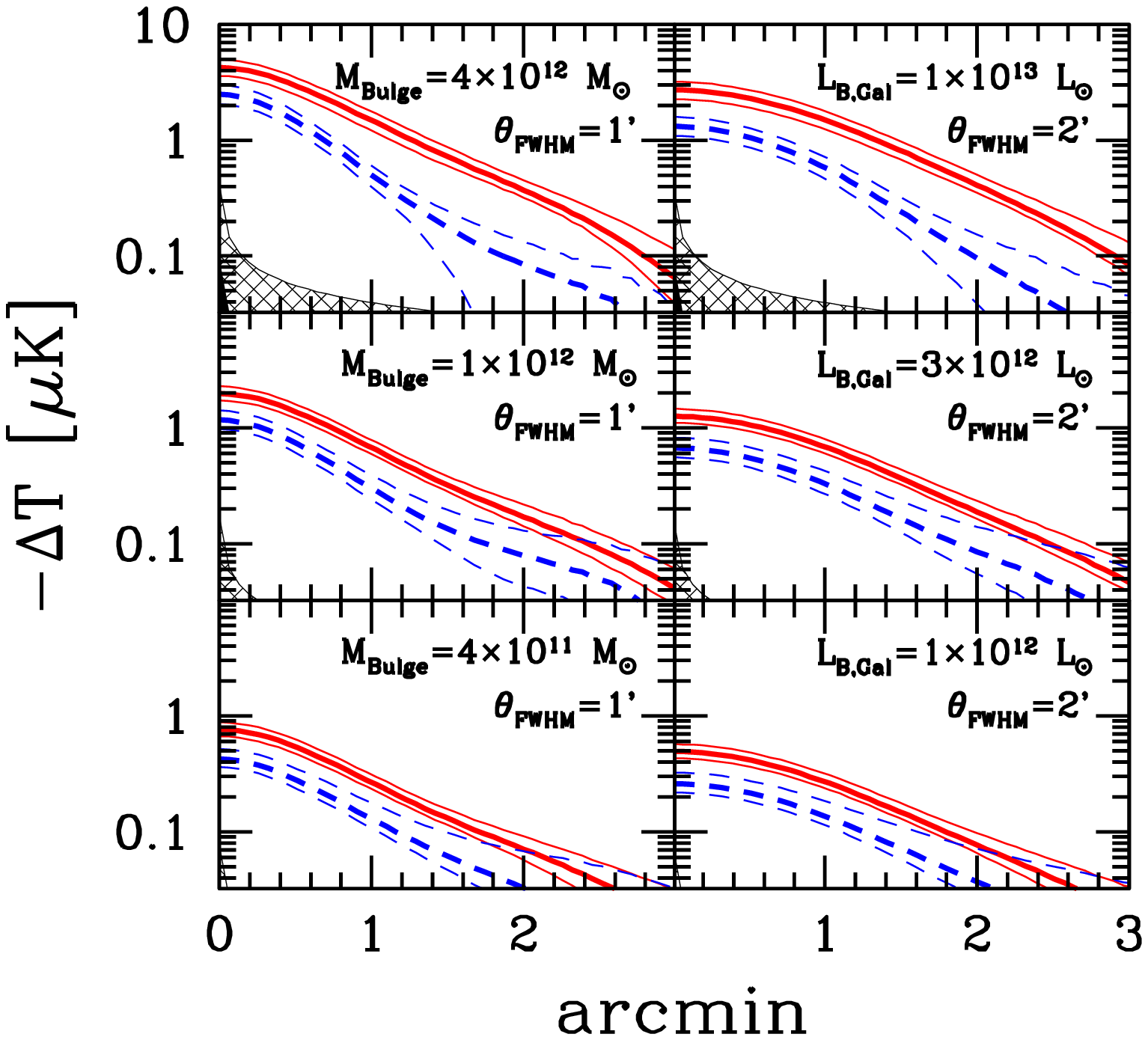,height=9.0cm}}
\caption{Average excess thermal SZ distortions around all bulges in
our simulations.  Mass limits, beam sizes, and symbols are as in
Figure \ref{fig:200}.  Right panels are labeled with estimated total
bulge $L_B$ luminosities following Marconi \& Hunt (2003)\\.}
\label{fig:1000}
\end{figure}

In Figure \ref{fig:1000} we show the SZ contribution obtained by
coadding all bulges observed at any time after the initial quasar
phase.  Again we have converted  black hole mass into bulge mass as
well as into  overall B-band luminosity following observed relations
(Marconi \& Hunt 2003).  Extending our analysis in this way has only a
weak impact on the SZ profile, while at the same time improving the
statistics even further.  In this  case the complete spatial profiles
of the sources are observable with more than 10 $\sigma$ precision out
to very large radii, clearly tracing out the  impact of AGN feedback.
Note that these sources are relatively easy to detect, as even given a
typical luminosity distance of $10^4$ Mpc, they are readily observed
by a photometric survey with an overall magnitude limit of $m_B
\approx 22.$

Furthermore, the level of thermal $y$ distortions observed by coadding
around bulges is  a precise and linear measure of the most important
quantity for modeling the impact of AGN outflows on galaxy formation.
Integrating eq.\ (\ref{eq:y}) over a patch of sky  around a source we
have 
\be 
\int d{\vec \theta} y({\vec \theta}) = \frac{\sigma_T}{m_e
c^2}  \frac{1}{l_{\rm ang}^2} \int dV n_e(V) k[T_e(V)-T_{\rm CMB}],
\label{eq:yenergy}
\ee 
where $l_{\rm ang}$ is the angular diameter distance to the
source, ${\vec \theta}$ is a vector in the plane of the sky in units of
radians,
and the volume integral is performed over the heated region
around the source.  But this integral is simply $(2/3) E_{\rm thermal}
(1+A)/(2+A)$ where $A = 0.08$ is the  cosmological number abundance of
helium.  
The line-of-sight integral of the pressure has thus been
transformed into a volume integral of the pressure.
This means that by integrating the Compton distortions over
the sky we can {\em directly measure} the thermal energy added to the
IGM for each object.  Rewriting eq.\ (\ref{eq:yenergy}) in terms of
angles in  arcminutes and the low-frequency microwave background
distortions in units  of $\mu K$ this becomes \be E_{\rm thermal} =
-4.8 \times 10^{60} \, {\rm ergs} \,\,\,\,  \tilde l_{\rm ang}^2
\frac{\int d{\vec \theta} \Delta  T({\vec \theta})}{\mu K \, {\rm
arcmin}^2},
\label{eq:Ethermal}
\ee where $\tilde l_{\rm ang}$  is the angular distance in units of
$3000$ Mpc and $E_{\rm thermal}$ is the total excess thermal energy
associated  with the source, that is the thermal energy gained from
the initial collapse of the baryons,  plus the contribution from the
AGN, minus losses due to cooling and $P dV$ work done during expansion.

\subsection{Measuring AGN Feedback}

Putting these results together, a clear and direct way to constrain
AGN feedback presents itself.  This can be summarized as follows:

{
\begin{itemize}
\item Obtain SZ data with 1 or 2 arcminute resolution at a noise level
of a few $\mu K$ per arcmin$^2$ in a $\approx 400$ deg$^2$ patch of the sky
in which  near infrared or  optical photometry is available to a
moderately deep magnitude limit (such as $m_B \approx 22$).

\item Select all quiescent early-type galaxies  with photometric
redshifts in  the essential $z=1-3$ redshift range for AGN feedback.  
Alternatively, if sufficiently good photometry is available, one can also 
study the subset of post-starburst  (E+A) galaxies.

\item If possible, use radio observations to reject point sources that
may contaminate the thermal SZ signal.

\item Compute the excess thermal SZ signal summed within a 
$\approx$ 2 arcminute radius 
 around each source, and use this to compute the
 total thermal energy as per eq.\  (\protect\ref{eq:Ethermal}).

\item Bin the results as a function stellar mass, and average over
each bin to compute the total average IGM thermal energy input
associated with bulges as function of their stellar mass.
\end{itemize}
}

\begin{figure}
\centerline{\psfig{figure=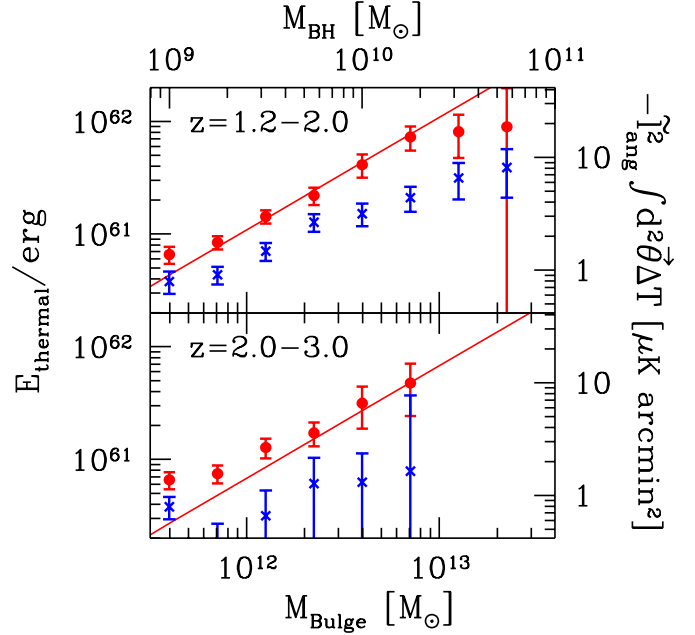,height=9.2cm}}
\caption{IGM thermal energy 
computed from eq.\ (\protect\ref{eq:Ethermal}) 
as a function of  bulge stellar mass 
in our AGN feedback (solid points) and
comparison (crosses) simulations.  In each bin the point shows the
results of averaging over all sources in a $\log_{10}(M_{\rm
bulge})=0.25$ bin surrounded by error bars that account for both the 
intrinsic variation between sources and the fact that multiple maps have been
derived from the same simulation  (see text).  The solid lines, on the
other hand, show the energy input added around bulges in our AGN
feedback simulation, as per eq.\  (\protect\ref{eq:ekin}).  In the
upper panel we consider all bulges in the redshift range from
$z= 1.2-2.0,$ and in the lower panel from $z=2.0-3.0.$\\}
\label{fig:Eplot}
\end{figure}

The results of carrying out this procedure over (2.2 deg)$^2$ of sky
calculated from our simulations are shown in Figure \ref{fig:Eplot}.
Here we have grouped all bulges in logarithmic bins of stellar mass with
width $0.25$, again taking a fixed ratio of 400 between $M_{\rm
bulge}$ and $M_{\rm BH}.$  To calculate the uncertainty in each bin we
have accounted both for Poisson noise and intrinsic scatter, 
taking $\sigma_{\rm bin}(E_{\rm thermal}) = [\bar E_{\rm thermal,bin} +
\sigma_{\rm source}(E_{\rm thermal})]/\sqrt{N_{\rm bin}},$ where $\bar
E_{\rm thermal,bin}$ is the average thermal energy in a given bin,
$\sigma_{\rm source}(E_{\rm thermal})$ is the RMS scatter between
sources in  a given bin, where we computed $N_{\rm bin}$ conservatively
as the number of sources in a given bin relative to a signal
realization of the map (1/4 of the [2.2 deg]$^2$ region in the AGN
feedback run and 1/16 of this region in the comparison run).

At both high and low redshifts the AGN feedback run shows a clear
excess, which scales linearly with $M_{\rm bulge}$ as expected from
eq.\ (\ref{eq:ekin}). For comparison, we use this equation to plot the
energy added to the IGM as a function of bulge mass.  Since kinetic
energy is converted into thermal energy as the outflows accrete
material, and as radiative losses are small for these objects (\eg Oh
\& Benson 2003; Scannapieco \& Oh 2004), we expect that most of this
energy should be observable as $E_{\rm thermal}.$  In fact, this
figure shows that at both low and high redshift, the thermal SZ excess
closely traces this energy input as a function of bulge mass.  Thus
AGN feedback is not only detectable by the SZ effect, but the level of
this feedback as a function of mass can be  obtained directly from SZ
measurements.

\begin{figure}
\centerline{\psfig{figure=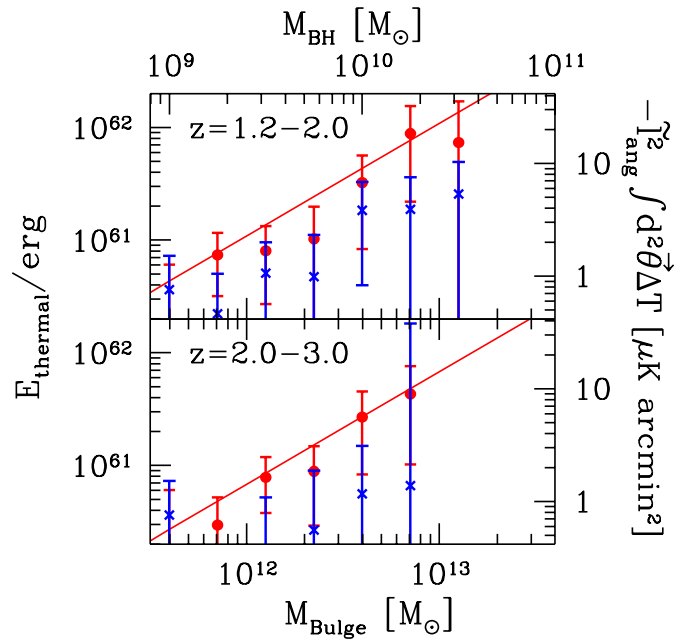,height=9.2cm}}
\caption{Thermal energy around post-starburst galaxies as a function of 
bulge stellar mass.   Panels and symbols are as in Figure
\protect\ref{fig:Eplot}.\\}
\label{fig:Eplot_E+A}
\end{figure}

Finally, in Figure \ref{fig:Eplot_E+A}, we plot $E_{\rm thermal}$ for
the  post-starburst galaxies in our simulations.   As there are fewer
of these objects, this results in a higher overall scatter than in the
case of all bulges, but the results are otherwise similar and
consistent with the level of excess energy added to the AGN feedback
run.  While this is to be expected given the instantaneous and
impulsive nature of the outflows in our simulation, comparisons of
post-starburst with other bulges may nevertheless help to distinguish
this type of feedback from more gradual AGN heating, as suggested, for
example, in Croton \etal (2006).

\section{Conclusions}

Microwave background measurements have played a key role throughout
cosmology: establishing the modern inflationary model, constraining
the initial conditions for structure formation, and providing precise
measurements of cosmological parameters.  Yet at smaller scales, the
potential of CMB  measurements is largely untapped.

At the same time, detailed observations at lower redshifts have
uncovered the central importance of nongravitational  heating, which
is likely to be associated with outflows from AGN.  Although the
details of this process remain uncertain, we have shown using a suite
of large numerical simulations that the key observations constraining
this  process may once again come from the microwave background.
While AGN outflows impact the CMB both through their peculiar motions
and through IGM heating, it is the thermal SZ signal that provides the
most detailed measurements.  Working purely with CMB observations, AGN
feedback can be constrained both through a difficult to measure
overall offset in the frequency spectrum, and a more easily detected
shift in CMB anisotropies from  $\approx 1/10$ arcmin to $\approx 1$
arcmin scales.

Although changes in the angular power spectrum from this shift are
marginally detectable with the upcoming generation of small-scale
microwave telescopes, the true test of AGN feedback will come from the
cross-correlation of this data with infrared and optical surveys.
In particular, coadding the CMB signal around quasars, post
starburst, and quiescent elliptical galaxies from our simulations
shows a systematic ``halo'' of SZ distortions around each of these
objects, which is much higher in amplitude and more spatially
extended in the AGN feedback case. Also, while contamination from
point source emission make quasars difficult to work with, the
excess signal is clearly detectable for both E+A and older elliptical
galaxies, which can be selected with only moderately deep $m_B \approx
22$ photometry.

Furthermore, as the thermal SZ effect is proportional to the line of
sight integral of the pressure, summing up the
excess  distortion in a patch of sky around each type of source
provides a sensitive and direct measure of the thermal energy
associated with feedback.  In fact, carrying out this procedure on our
simulation we have shown that we can easily recover not only the
overall level of AGN feedback, but its dependence on galaxy mass,
redshift, and galaxy type.   Although it provides a popular and
elegant solution to many  outstanding problems, the impact of IGM heating
by AGN outflows remains perhaps the most uncertain outstanding issue
in galaxy formation.  Working together with large-field optical
surveys, small-scale CMB experiments will soon be  able to place
strong constraints on this missing piece in our physical
understanding of the history of our universe.

\acknowledgments

ES was supported by the National Science Foundation under grants
PHY99-07949 and AST02-05956 to the Kavli Institute for Theoretical
Physics, where this work was initiated.  RJT acknowledges funding via
a  Discovery Grant from NSERC, the  Canada Foundation for Innovation
and  Canada Research Chairs programme.  Simulations and analysis were
conducted on WestGrid facilities at the  University of Alberta and
Simon Fraser University.  HMPC acknowledges funding  from NSERC and
the support of the Canadian Institute for Advanced Research.

\fontsize{10}{10pt}\selectfont

\end{document}